\documentclass[12pt]{article}
\usepackage{amsmath,amssymb,color}
\usepackage{cite,epsf}
\usepackage{pstricks}
\usepackage{color}
\usepackage{graphicx,array} 
\newcommand{\be}{\begin{equation}}
\newcommand{\ee}{\end{equation}}
\newcommand{\beq}{\begin{equation}}
\newcommand{\eeq}{\end{equation}}
\newcommand{\bea}{\begin{eqnarray}}
\newcommand{\eea}{\end{eqnarray}}

\newcommand{\ba}{\begin{eqnarray}}
\newcommand{\ea}{\end{eqnarray}}

\begin{document}

\begin{titlepage}
\vspace{10pt}
\hfill
{\large\bf HU-EP-11/54}
\vspace{20mm}
\begin{center}

{\Large\bf  Hexagon remainder function in the limit \\[2mm] 
of self-crossing up to three loops}

\vspace{45pt}

{\large Harald Dorn, Sebastian Wuttke
{\footnote{dorn@physik.hu-berlin.de, wuttke@physik.hu-berlin.de
 }}}
\\[15mm]
{\it\ Institut f\"ur Physik der
Humboldt-Universit\"at zu Berlin,}\\
{\it Newtonstra{\ss}e 15, D-12489 Berlin, Germany}\\[4mm]

\vspace{20pt}

\end{center}
\vspace{10pt}
\vspace{40pt}

\centerline{{\bf{Abstract}}}
\vspace*{5mm}
\noindent
We consider Wilson loops in planar ${\cal N}=4$ SYM for null polygons
in the limit of two crossing edges. The analysis is based on a 
renormalisation group technique. We show that the previously obtained
result for the leading and next-leading divergent term of the two loop
hexagon remainder is in full agreement with the appropriate continuation
of the exact analytic formula for this quantity. Furthermore, we 
discuss the coefficients of the leading and next-leading singularity
for the three loop remainder function for null $n$-gons with $n\geq 6$.
   
\end{titlepage}
\newpage

%\tableofcontents \newpage

\section{Introduction}
Recently, Wilson loops for polygonal null contours in planar ${\cal N}=4$ 
supersymmetric Yang-Mills theory have attracted a lot of attention. This interest 
is mainly due to their correspondence to gluon scattering amplitudes. It has been 
first established at strong coupling by string theoretical arguments 
\cite{alday-malda} and later verified also at weak coupling \cite{drummond}. 
On the perturbative side one expresses the logarithm of the Wilson loop as the
sum of the well-known BDS structure \cite{bds} plus some remainder function.
Then the determination of this  remainder function in higher and higher loop 
order comes into the main focus.

At two-loop order the complete hexagon remainder \cite{duhr,goncharov} and
the octagon remainder for restricted configurations \cite{duhr-8} are available.
Furthermore, the differential of the remainder has been analysed for arbitrary
null $n$-gons yielding explicit analytic formul\ae \, for some parts while only giving the symbol 
for other terms \cite{caron}. At three-loop level
a recent paper has bootstrapped the symbol of the hexagon remainder \cite{henn}.
The results of \cite{caron,henn} have been obtained by combining symmetry arguments
and constraints provided by  operator product expansions and by collinear
as well as by multi Regge limits. 

Further independent information on the remainder functions can be
extracted out of the study of the limit of self-crossing contours. Such an analysis
can be based on standard renormalisation (RG) group techniques\footnote{We would like to stress that this technique can give results
also for QCD. There the high symmetry being crucial for the bootstrap programs
is not available.}. 
It has been started
for the two-loop hexagon remainder in \cite{georgiou} and continued 
in our previous paper on the two-loop octagon case \cite{dw}. To be more precise,
the difference between the two papers is not so much an issue of hexagon versus
octagon. A self-crossing polygon can be one with crossing edges or one with
two coinciding vertices. For a hexagon, due to the null condition, only the
first case can be realized, for octagons and higher $n$-gons both cases are possible.
In the limit with crossing edges the cross-ratio $u$ formed out of the
four endpoints of the edges approaches the number one, and the remainder develops 
singularities in powers of $\log(u-1)$ with pure numerical coefficients. In the 
second type of limit one gets divergences in powers of the logarithm of a product 
of four cross-ratios.  This time the coefficients depend on the cross-ratio 
formed out of the four vertices adjacent to the two coinciding ones.  

The aim of this paper is twofold. First we want to fill a gap left by 
\cite{georgiou,dw}. As described in detail in these papers, the goal to
control the short distances singularities of the remainder function
in the approach to a self-crossing configuration is achieved in the following way. 
Using  dimensional regularisation, one analyses the additional UV-singularities 
due to the crossing directly in a self-crossing configuration.
Then one interprets the near to self-crossing situations as some alternative
regularisation, providing a translation of powers in $1/\epsilon$
into powers of the negative logarithm of some distance. Due to the usual
scheme ambiguities, in this translation rule numerical factors remain open.
We want to fix these numbers with arguments based on our approach \emph{before}
we show the agreement of the two-loop hexagon result of \cite{georgiou,dw} with
the appropriate analytic continuation of the complete remainder function of
\cite{goncharov}.

The second task concerns the calculation of the leading plus next-leading
divergent terms for the three-loop hexagon (or even $n$-gon) remainder function for the limit of two edges approaching  the self-crossing configuration. 
%%%%%%%%%%%%%%%%%%%%%%%%%%%%
\section{RG technique for the remainder function at two and three loop level}
\begin{figure} 
 \centering
 \includegraphics[width=6cm]{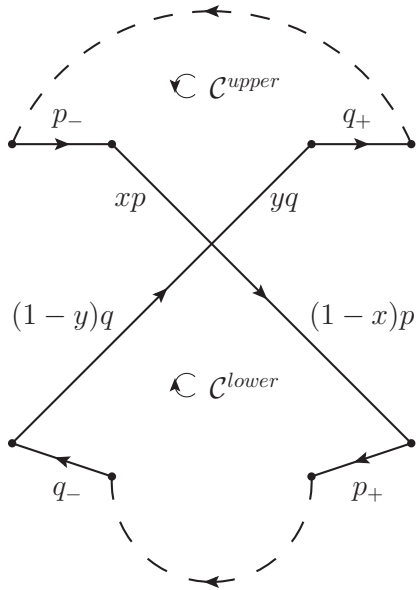}
 % Cross.eps: 0x0 pixel, 300dpi, 0.00x0.00 cm, bb=189 432 406 721
 \caption{{\it A self-crossing Wilson loop with  crossing edges  $p$ and $q$. The dashed lines represent any light-like continuation. The dimensionless parameters $x,y \in (0,1)$ define where the crossing is located exactly.}}
\label{Cross}
\end{figure}

Self-crossing Wilson loops were first studied for QCD  and it was shown  \cite{pol,brandt,dorn,korchemsky2}   that they can be multiplicatively renormalised using a $\mathcal Z$ matrix
\begin{equation}
\mathcal W_a ~=~ \mathcal Z_{ab} ~ \mathcal W_b^{\text{ren}}~.\label{WZW}
\end{equation}
$\{\mathcal W_a\}$ is the set of mixing Wilson loop operators,  $a\in \{1,2\}$.\\
With   $\mathcal U(\mathcal C ) :=  \frac{1}{N} ~ \text{tr}~ \mathcal P ~\exp \left(   ig  \int_{\mathcal C}  A^\mu \text{d}x_\mu \right)$ for $\text{SU}(N)$ theory. $\mathcal W_1$ and $\mathcal W_2$ are defined by
\begin{equation}
\mathcal W_1   := \langle \mathcal U(\mathcal C) \rangle~,~~\mathcal W_2   := \langle \mathcal U(\mathcal C^{\text{upper}}) \mathcal U(\mathcal C^{\text{lower}})\rangle ~.
\label{wilson}
\end{equation}
In our case all contours $\mathcal C^{\text{upper}}$, $\mathcal C^{\text{lower}}$ and $\mathcal C$ are light-like closed polygons. $\mathcal C$ is the full $n$-cusps self-crossing contour (see fig.\ref{Cross}) and $\mathcal C^{\text{upper}}$ and $\mathcal C^{\text{lower}}$ are the two sub-contours starting and ending at the crossing point.

The vertices of the polygons are denoted by $x_i$, and because of the correspondence to scattering amplitudes we call $p_i = (x_{i+1}-x_i)$ a momentum. The two crossing edges (momenta) are denoted by $p$ and $q$, and  the fractions that determine the crossing point by  $x$ and $y$ (see fig.\ref{Cross}). In  the 't Hooft limit  under consideration, $ \mathcal W_2$ factorises 
\begin{equation}
 \mathcal W_2 = \langle \mathcal U(\mathcal C^{\text{upper}}) \rangle  \langle 
\mathcal U(\mathcal C^{\text{lower}}) \rangle ~,
\end{equation}
and the  $\mathcal Z$ -matrix has the triangular form 
\begin{equation}
\mathcal Z = \begin{pmatrix}\mathcal Z_{11} & \mathcal Z_{12} \\ 0 & \mathcal Z_{22}  \end{pmatrix}~.\label{z-matrix}
\end{equation}  
Then also the anomalous dimension matrix
\begin{equation}\label{defGamma}
\Gamma := \mathcal Z^{-1} \mu \frac{d}{d\mu} \mathcal Z~\Big \vert _{g_{\text{bare}}{\text{~fixed}}}~,
\end{equation}
 is upper triangular. 

Originally, the renormalisation of Wilson loops has been studied for contours
without light-like pieces. Then the $\mathcal Z$-factors as well as the anomalous dimensions
depend on the coupling constant and the angles at the cusps and at the self-crossing.
In the light-like limit  the angles become divergent and make the related 
original RG equation ill-defined. Following  \cite{korchemsky} one nevertheless
can get a modified RG equation which has
the same structure as the original one, but with RG scale dependent anomalous
dimensions. For more details and references we refer to \cite{korchemsky},\cite{korchemsky2},\cite{georgiou,dw}.

The RG equation to start with is then 
\footnote{
There is no derivative with respect to the coupling since the $\beta$-function in ${\cal N}=4$ SYM is zero in four dimensions. }
$
\mu   \frac{ \partial }{\partial\mu}  ~ \mathcal W_a^{\text{ren}}~ =~ -~ \Gamma_{ab}~\mathcal W_b^{\text{ren}}~,~ a,b \in \{1,2\}~.$
It implies for $\mathcal W_1^{\text{ren}}$
\begin{equation}\label{RGE1}
\mu  \frac{\partial}{\partial\mu}   ~\log \mathcal W_{1}^{\text{ren}}~ = ~-~ \Gamma_{12}~\frac{\mathcal W_2^{\text{ren}}}{\mathcal W_1^{\text{ren}}}~-~\Gamma_{11}~.
\end{equation}
For the following analysis we need formul\ae \, expressing $\log \mathcal W_{a}^{\text{ren}}$ in terms of its unrenormalised partners and the entries of the anomalous dimension matrix. Preparing this for the use up to three loop level requires a bit more
effort as in \cite{georgiou,dw}. A further comment concerns the fact, that for self-crossing polygons with crossing edges,  the  $\mathcal Z$-factors no longer factorise into a cusp and a crossing part. Therefore,  one has  to
look for the relation of the overall $\mathcal Z$-matrix in \eqref{WZW} and the total
$\Gamma$-matrix in \eqref{RGE1}.

The anomalous dimension of a Wilson loop for a self-crossing null polygon is given \cite{korchemsky},\cite{korchemsky2},\cite{georgiou} via
\begin{equation}
\Gamma = ~\begin{pmatrix} 1 & 0\\
0 &1 \end{pmatrix} ~ \frac{\Gamma_{\text{cusp}}(a)}{2}\sum_{k\in \text{cusps}} \log\left(-s_k \mu^2 \right) +~\begin{pmatrix} A & \gamma_{12}(a)\\
0 &B \end{pmatrix} ~,\label{gamma}
\end{equation}
where
\begin{align}
A =& \frac{\Gamma_{\text{cusp}}(a)}{2}\left(\log (-2pp_{-} \mu^2 )+\log (-2pp_{+} \mu^2 )+\log (-2qq_{-} \mu^2 )+\log (-2qq_{+} \mu^2 ) \right)~, \nonumber\\
B =& \frac{\Gamma_{\text{cusp}}(a)}{2}\left( \log (-2pp_{-} x \mu^2 )+\log (-2pp_{+} (1-x) \mu^2 )
+\log (-2qq_{-} (1-y) \mu^2 )\right .\nonumber\\
&\left . +\log (-2qq_{+} y \mu^2 ) \right)+\gamma_{22}(a)\left(\log(-sxy\mu^2 )+\log(-s(1-x)(1-y)\mu^2 )\right) \nonumber ~,
\end{align}
with $a=\frac{g^2 N}{8 \pi^2}$ and $s=2pq$. By $\sum_{k\in \text{cusps}}$ we denote all cusps that are {\emph{not}} adjacent to the crossing and by $s_k=(x_{k-1}-x_{k+1})^2$ the Mandelstam variables at the cusps. 

All entries in the anomalous dimension matrix that are proportional to $\Gamma_{\text{cusp}}$ originate from UV divergences that are related to cusps. The cusps adjacent to the crossing contribute in a different manner to $\mathcal W_1$ and $\mathcal W_2$. For $\mathcal W_1$ the full momenta $p$ and $q$ are relevant and for $\mathcal W_2$ only fractions thereof. Therefore, also the cusp terms in the anomalous dimension matrix are \textit{not} a multiple of the unit matrix. In the anomalous dimension matrix $\gamma_{12}$ and $\gamma_{22}$ are functions of the coupling only. The one-loop results in planar limit are 
\begin{equation}
\gamma_{12}^{(1)}~= ~\pm ~\text{sgn}(pq)~ 2 \pi i~~,~~~~\gamma_{22}^{(1)}~=~1~~,~~~~
\Gamma^{(1)}_{\text{cusp}}~=~2~.
\end{equation}
We will later comment on $\gamma_{12}^{(1)}$ as this entry plays an important role for the analytic continuations discussed in the next section.

Now we consider \eqref{defGamma} as a differential equation in $\mu$  for $\mathcal Z$. Because of the upper triangular structure of $\mathcal Z$ and $\Gamma$ we find
\begin{align}
\mu \frac{d}{d\mu} \log \mathcal Z_{11} ~&= \Gamma_{11}~,~~~~~~~~
\mu \frac{d}{d\mu} \mathcal Z_{12}=\mathcal Z_{11}\Gamma_{12} + \mathcal Z_{12} \Gamma_{22}  \label{eqZ12}~,  \\
\mu \frac{d}{d\mu} \log \mathcal Z_{22}~&=\Gamma_{22}\nonumber~.
\end{align}
We solve this as an expansion in powers of $a$, i.e. $\Gamma =\sum_l a^l\Gamma^{(l)}$ and similar for $\mathcal Z$. Keeping in mind that the $\mu$ derivative has to be taken at fixed bare coupling and that $a = a_{\text{bare}}\mu^{-2\epsilon}$, the first two equations give for a $n$-gon 
\begin{align}
\mathcal Z_{11}^{(0)}&=1~,~~~~~
\mathcal Z_{11}^{(1)}~=-\frac{n \Gamma^{(1)}_{\text{cusp}}}{4\epsilon^2} -\frac{\Gamma_{11}^{(1)}}{2\epsilon}~,\nonumber\\
\mathcal Z_{22}^{(0)}&=1~,~~~~~
\mathcal Z_{22}^{(1)}~=-\frac{n \Gamma^{(1)}_{\text{cusp}}+4\gamma_{22}^{(1)}}{4\epsilon^2} -\frac{\Gamma_{22}^{(1)}}{2\epsilon}~.
\end{align}
Now we solve \eqref{eqZ12} for $\mathcal Z_{12}$. A priori it is clear that $\mathcal Z_{12}^{(0)}=0$.  At order $\mathcal O(a)$ we find
\begin{equation}
\Gamma^{(1)}_{12} a = \mu \frac{d}{d\mu}\left( \mathcal Z_{12}^{(1)} a \right) ~~ \Longrightarrow~~ \mathcal Z_{12}^{(1)} = -\frac{\gamma_{12}^{(1)}}{2 \epsilon}~.\label{z12}
\end{equation}
At order $\mathcal O(a^2)$ eq.\eqref{eqZ12} means
\begin{equation}
a^2\left( \Gamma_{12}^{(2)} + \mathcal Z_{11}^{(1)} \Gamma_{12}^{(1)} + \mathcal Z_{12}^{(1)} \Gamma_{22}^{(1)}  \right) = \mu \frac{d}{d \mu}\left( \mathcal Z_{12}^{(2)}a^2\right)~.
\end{equation}
Integration yields
\begin{equation}
\mathcal Z_{12}^{(2)} = \frac{( n\Gamma^{(1)}_{\text{cusp}} +\gamma_{22}^{(1)})  \gamma_{12}^{(1)}}{8 \epsilon^3}+ \frac{\gamma_{12}^{(1)}\left( \Gamma_{11}^{(1)}+\Gamma_{22}^{(1)} \right)}{8 \epsilon^2}-\frac{\gamma_{12}^{(2)}}{4\epsilon}~.
\end{equation}
Expanding the logarithms of the Wilson loops in powers of $a$ one gets from \eqref{WZW} within
minimal subtracted dimensional regularisation  
\begin{align}
\log \mathcal W_{1}^{\text{ren}(1)} &= \text{MS} \Big [ \log \mathcal W_{1}^{(1)} \Big ] = \mathcal W_{1}^{\text{ren}(1)}~,\nonumber \\
\log \mathcal W_{1}^{\text{ren}(2)} &= \text{MS} \Big [ \log \mathcal W_{1}^{(2)}  + \mathcal Z_{12}^{(1)} \left( \mathcal W_1^{\text{ren}(1)}-\mathcal W_2^{\text{ren}(1)} \right) \Big ]~,\label{defT}\\
\log \mathcal W_{1}^{\text{ren}(3)} &= \text{MS} \Big [ \log \mathcal W_{1}^{(3)} ~ - ~T_1 ~ - ~T_2   \Big ] ~, \nonumber
\end{align}
with  $\text{MS}\big[ ~. ~\big]$ denoting minimal subtraction and
\begin{align}
T_1 &:= \mathcal Z_{12}^{(1)} \left( \frac{1}{2} \left( \mathcal W_1^{\text{ren}(1)}-\mathcal W_2^{\text{ren}(1)} \right)^2 - \log \mathcal W_{1}^{\text{ren}(2)} +\log \mathcal W_{2}^{\text{ren}(2)} \right)~,\label{Tdef1} \\
T_2 &:=\left(\left( \mathcal Z_{12}^{(1)}  \right)^2 +\mathcal Z_{12}^{(1)} \mathcal Z_{11}^{(1)} - \mathcal Z_{12}^{(2)} \right)\left( \mathcal W_1^{\text{ren}(1)}-\mathcal W_2^{\text{ren}(1)} \right)~. \label{Tdef2}
\end{align}
In contrast, due to the triangular form of the $\mathcal Z$-matrix in \eqref{z-matrix}, for $\mathcal W_2$ one simply has
\begin{equation}
\log \mathcal W_{2}^{\text{ren}(l)}~ = ~ \text{MS} \Big [~ \log  \mathcal W_{2}^{(l)} ~ \Big ]~.
\end{equation}
For the logarithms of the Wilson loops we insert the BDS Ansatz \cite{bds},  corrected by the remainder function $\mathcal R$~.
\begin{equation}
\log \mathcal W = \sum_{l=1}^\infty a^l \left( f^{(l)}(\epsilon) w(l~\epsilon)+C^{(l)} \right) + \mathcal R + \mathcal O(\epsilon)~,\label{logw}
\end{equation}
where $C^{(l)}$ are (known) numbers, $f^{(l)}(\epsilon) = f_0^{(l)}+\epsilon~f_1^{(l)}+\epsilon^2~ f_2^{(l)}$, and $w(\epsilon)$ is the one loop contribution
\begin{equation}
w(\epsilon) ~=~ - \frac 1 {2 \epsilon^2} \sum_{k=1}^n \left( - \mu^2 s_k \pm i \varepsilon \right)^\epsilon + F(\mu^2,\epsilon,\{s\})~.
\end{equation}
Here $n$ is again the number of cusps and $s_k$ the Mandelstam variable associated with the $k$'th cusp\footnote{We use $\epsilon$ as the regulator in dimensional regularisation. In contrast, $\varepsilon$ is used for the $i \varepsilon$-pole prescription originating from the gluon propagator. The reason for keeping an option  $\pm$ will be explained in the next section.}. 
$F(\mu^2,\epsilon,\{s\})$ is the so-called one-loop finite contribution. It remains 
finite for $\epsilon \rightarrow 0$  in a generic $n$-gon configuration, but develops
poles in a self-crossing case. The $a$-expansion of $\mathcal R$ starts at $\mathcal O(a^2)$.

It will be convenient  to include the $\mathcal O(\epsilon)$ term in \eqref{logw} into
$\mathcal R$ i.e. to use 
\begin{equation} \label{defR}
\mathcal R(\mu^2,\epsilon, \{s\}) = \log \mathcal W - \Big[ \text{BDS}\Big]~,
\end{equation}
where $\{s\}$ is the set of all Mandelstam variables. For a generic Wilson loop 
\begin{equation}
\mathcal R(\{u\}) = \lim_{\epsilon \rightarrow 0} ~ \mathcal R(\mu^2,\epsilon, \{s\})
\end{equation}
is finite, independent of $\mu^2$ and only a function of conformal invariants $\{u\}$. However, if the Wilson loop contains a self-crossing the new divergences at the crossing point will lead to poles in $\epsilon$ in $\mathcal R(\mu^2,\epsilon, \{s\})$. 

Our aim is to calculate the (most) divergent terms. Plugging the BDS Ansatz including  the remainder function into the renormalisation group equation \eqref{RGE1} one can calculate $\text{MS} \Big[ \mathcal R(\mu^2,\epsilon, \{s\}) \Big]$ using \eqref{defT},\eqref{defR}. From this expression one can deduce the most divergent terms, as we will do explicitly for the three-loop remainder in section 4. Then, considering a slight off self-crossing configuration as an alternative regularisation, one can relate those divergent terms to divergences of $\mathcal R(\{u\})$ when the cross-ratios approach their self-crossing values. This method works for the leading and next-leading divergences at all orders in $a$. To calculate those terms, only one-loop information on the anomalous dimension matrix is needed. The leading and next-leading contributions for $\mathcal R^{(2)}$ were calculated in \cite{georgiou}\footnote{There has been
forgotten a contribution due to renormalisation $Z$-factors. Taking it into account leads to an overall factor of two, as
has been observed in our previous paper \cite{dw}.}. When this paper was written, the compact analytic result of \cite{goncharov} was not yet available. 
The remainder function presented in \cite{goncharov} is finite for these  values of the cross ratios. However, the remainder function is multi-valued. Wilson loops for null hexagons with all Mandelstam variables in the Euclidean region can never be conformal to self-intersecting ones. The divergent terms are created by the analytic continuation to a self-crossing region, as we will explain in the next section. 
%%%%%%%%%%%%%%
\section{Analytic continuation from the Euclidean region}
%\noindent
\begin{figure}
 \centering
 \includegraphics[width=12cm]{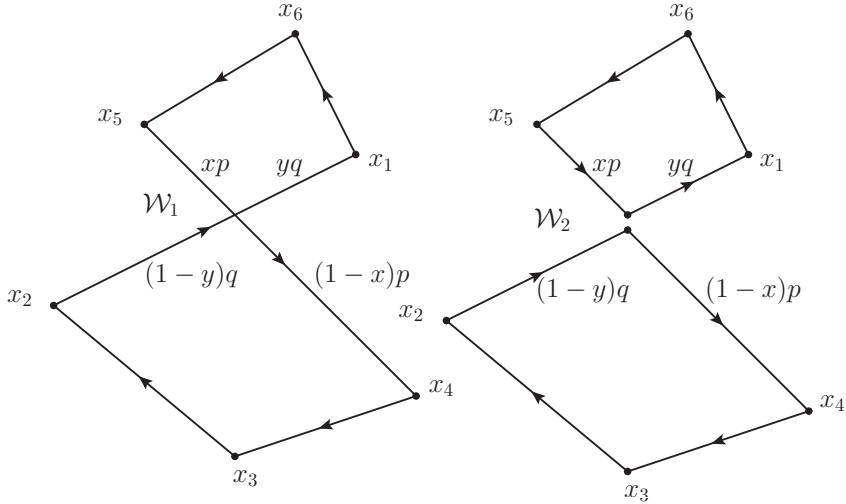}
 % CrossHexa.eps: 0x0 pixel, 300dpi, 0.00x0.00 cm, bb=71 470 583 720
 \caption{\it  The self-crossing hexagon and its mixing partner. }
 \label{HexaCross}
\end{figure} 
For the comparison of the asymptotic result of \cite{georgiou} with certain analytic continuations of the exact hexagon remainder of \cite{goncharov} one has to be very careful with the $i\varepsilon$-prescription and the induced signs of imaginary parts.

With respect to the correspondence
between Wilson loops and scattering amplitudes it has been argued, that
for the use in this duality one has to switch the sign of the $i\varepsilon$-prescription in the gluon propagator relative to that in the  standard position space 
propagator \cite{georgiou},\cite{brandhuber10}. This switch leads to a change of the sign of the one loop entry 
$\gamma_{12}^{(1)}$ of 
the crossing anomalous dimension matrix \cite{georgiou} in comparison
to the original calculation of \cite{korchemsky2}.
 
Denoting by $p$ and $q$ the vectors spanning the crossing edges of the
hexagon as in fig.\ref{HexaCross} one gets according to \cite{korchemsky2}
\bea
pq<0&:&\gamma_{12}^{(1)}=-2\pi i\coth \beta ~,~~~~\cosh\beta =-\frac{pq}{\sqrt{p^2q^2}}\nonumber\\
pq>0&:&\gamma_{12}^{(1)}=2\pi i\coth \gamma ~,~~~~\cosh\gamma=\frac{pq}{\sqrt{p^2q^2}}~.\label{gamma-korch-0}
\eea
Keeping now in addition the option of both signs in the $i\varepsilon$-prescription, we
get in the light-like limits ($\beta\rightarrow\infty$, $\gamma\rightarrow\infty$)
\beq
\gamma^{(1)}_{12}~=~\pm~\mbox{sgn}(pq)~2\pi i\label{gamma-korch}~.
\eeq
Here the upper alternative stands for the standard choice  
and the lower one for Georgiou's switched version. Let us also introduce the following abbreviation
\begin{equation}
\mathcal X := x~y~(1-x)~(1-y)~~,~~~~x,y \in (0,1) ~.\label{X}
\end{equation}

After this preparation we get the remainder function in a self-crossing configuration for $pq<0$ as
\beq
\mathcal R^{(2)}=\mp ~\frac{\pi i}{4}\left (\epsilon ^{-3}+2 \log \left (|2pq|\mu^2 \mathcal X  \right )~\epsilon^{-2}\right ) +\frac{\pi^2}{2}~\epsilon^{-2}+{\cal O}(\epsilon^{-1})~,
\label{2-loop-rem}
\eeq
and for $pq>0$ as
\beq
\mathcal R^{(2)}=\pm~\frac{\pi i}{4}\left (\epsilon ^{-3}+2 \log \left (2pq\mu^2 \mathcal X  )\right )~\epsilon^{-2} \right )~~+~~{\cal O}(\epsilon^{-1})~.
\label{2-loop-rem,2}
\eeq
The light-like closed hexagon has three conformal invariants $u_1$, $u_2$ and $u_3$, where
\begin{equation}
u_1 := \frac{x_{13}^2 ~ x_{46}^2}{x_{14}^2~ x_{36}^2}~,~~u_2 := \frac{x_{15}^2 ~ x_{24}^2}{x_{14}^2~ x_{25}^2}~,~~u_3 := \frac{x_{35}^2 ~ x_{26}^2}{x_{25}^2~ x_{36}^2}~,
\end{equation}
and $x_{ij}^2 = (x_i -x_j)^2$. For self-crossing hexagons the kinematics are restricted such that one cross ratio equals one and the remaining two cross ratios are equal to each other. We follow the choice of \cite{georgiou} and use $u_2=1$.

We now want to translate the information \eqref{2-loop-rem},\eqref{2-loop-rem,2} into a statement on the remainder 
function
in the limit where one approaches a self-crossing situation with $u_2=1$, coming from a generic configuration without self-crossing. Denoting by $z$ the vector between 
the points marked by the fractions $x$ and $y$ on the edges $p$ and $q$, 
respectively, one gets (see fig.\ref{HexaCross})
\beq
u_2-1~=~\frac{2pq~z^2-4(qz)(pz)}{(-2y(1-x)pq +{\cal O}(z))(-2x(1-y)pq+{\cal O}(z))}
~.\label{u-z}
\eeq 
Writing the arbitrary vector $z$ as $z=\alpha p+\beta q+z_{\perp}$ with $pz_{\perp}=qz_{\perp}=0,~z_{\perp}^2\leq 0$, the nominator of \eqref{u-z} turns out to be
equal to $2pq~z_{\perp}^2$. This implies that for $pq<0$ \\( $pq>0$ ) the approach to a self-crossing situation is possible only with $u_2\rightarrow 1$ from above ( below ). 
For $pq<0$ we get from \eqref{u-z}
\beq
\log \Big (\frac{1}{-z_{\perp}^2\mu^2}\Big )~=~-\log(u_2-1)~-~\log \big(-2pq\mu^2 \mathcal X \big )~+~{\cal O}(z^2)~.\label{log-u-z}
\eeq
For $pq>0$ it is more convenient to use instead
\beq
\log \Big (\frac{1}{-z_{\perp}^2\mu^2}\Big )~=~-\log(1-u_2)~-~\log \big(2pq\mu^2 \mathcal X \big )~+~{\cal O}(z^2)~.\label{log-u-z-2}
\eeq 
Then all terms in \eqref{log-u-z} and \eqref{log-u-z-2} are real.
While in the limit under consideration in both equations the l.h.s and the first term of the r.h.s. diverge, the second r.h.s. term stays finite.

Considering dimensional regularisation and point splitting via the introduction
of the separation vector $z$ as two regularisations of the UV divergences due
to the self-crossing, one expects a translation rule of the type
\beq
\frac{1}{\epsilon ^m}~\Leftrightarrow ~\alpha_m ~\log^m\Big(\frac{1}{-z_{\perp}^2\mu^2}\Big)~.\label{2-loop-trans}
\eeq
In appendix B we derive, with some heuristic arguments, a general translation rule
for arbitrary loop order. For the two-loop case it implies $\alpha_3=2/3$ and $\alpha_2=1$. This yields for $pq<0$
\beq
\mathcal R^{(2)}~=~\pm~\frac{i\pi}{6}\log ^3(u_2-1)~+~\frac{\pi^2}{2}\log^2(u_2-1)~+~{\cal O}\big (\log(u_2-1)\big )~,\label{R-hexa}
\eeq 
and for $pq>0$
\beq
\mathcal R^{(2)}~=~\mp~\frac{i\pi}{6}\log ^3(1-u_2)~~+
~~{\cal O}\big (\log(1-u_2)\big )~.\label{R-hexa-2}
\eeq 
At this stage we want to stress an observation, which gives additional
support to our translation rule from appendix B. The relative weight
$\alpha_3/\alpha_2$ is just the right one to prevent the appearance
of the not conformally invariant part of the coefficient of $1/\epsilon^2$ 
in \eqref{2-loop-rem},\eqref{2-loop-rem,2}
 in the final result for $\mathcal R^{(2)}~$! 

The complete analytic result of \cite{goncharov} has the form
\beq
\mathcal R^{(2)}~=~-\frac{1}{2}\mbox{Li}_4(1-1/u_2)~-~\frac{1}{8}\big( \mbox{Li}_2(1-1/u_2)\big )^2~+~\dots ~.\label{R-gonch}
\eeq 
It has been derived in the Euclidean region where all distances between non
adjacent vertices of the hexagon are space-like. The dots in \eqref{R-gonch}
stand for terms which for sure are not a source for a divergent contribution
in our self-crossing limit. Staying in the first sheet of the Riemann surface
of the polylogarithms, which is defined by the Euclidean region, one hits no 
singularity at $u_2=1$, i.e. $v:=1-1/u_2=0$. However, there is a singularity
at $v=0$ in the second sheet, which can be reached via encircling the branch point
at  $v=1$, which is the starting point for a cut extending along the real axis
from $1$ to $+\infty$. It is due to the continuation of the discontinuity 
across the cut 
\beq
\mbox{Li}_n(v+i \varepsilon)-\mbox{Li}_n(v-i \varepsilon)=\frac{2\pi i\log ^{n-1}v}{(n-1)!}~.\label{disc}
\eeq 
What concerns real hexagon configurations, for $pq< 0$ one approaches $v=0$ from
above and the logarithms stay real. For $pq>0$ the approach proceeds from
negative values of $v$ and the sign of the imaginary part matters.

A starting point in the Euclidean region necessarily has $u_2>0$ i.e. $v<1$. Encircling $v=1$ in a counterclockwise (clockwise) manner and approaching then $v=0$ from above gives just the standard (switched) alternative in \eqref{R-hexa}.

Encircling $v=1$ in a clockwise (counterclockwise) manner, going then to
negative values of $v$ via a detour around $v=0$ with the same orientation
and the approach to $v=0$ from below gives the standard (switched) alternative in \eqref{R-hexa-2}.

To finish the proof that the complete analytic result of \cite{goncharov} implies
our formul\ae ~ for the approach to a self-crossing configuration, one still has to
clarify a subtlety. There is no one to one correspondence between
the points in the space of cross-ratios and the classes of conformally equivalent
hexagon configurations. Therefore, the above identification of suitable paths for 
analytic continuation in the cross-ratios still does not necessarily mean that
these paths can be generated by smooth deformations of the hexagon from the 
Euclidean region. To fill this gap, we construct in appendix \nolinebreak A an 
explicit
example for a smooth deformation of a configuration in the Euclidean region
to a self-crossing configuration with $pq<0$. In this case $1/u_2$ goes from
one to zero and back to one. Combining the ``reflection'' at $1/u_2=0$ with
an encircling, we just get the paths in $z=1-1/u_2$ described above in connection
with \eqref{R-hexa}. Even more, the orientation turns out to be
correct if the encircling is generated by introducing an $i\varepsilon$-prescription in the cross-ratios by replacing the factors $(x_j-x_k)^2$ by $(x_j-x_k)^2\mp i\varepsilon$ just in the same manner as in the gluon propagator.          
%%%%%%%%%%%%%%%%%%%%%%%%%%%

\section{Three-loop remainder function}

We are interested in the leading and next-leading divergences of the remainder function of a $n$-sided light-like closed Wilson loop (for $n \geq 6$) in $\mathcal N=4$ SYM in the limit of two crossing edges.
Expanding the integrated version of \eqref{RGE1} in powers of $a$ one gets
with \eqref{defT},\eqref{Tdef1},\eqref{Tdef2},\eqref{defR}
\begin{equation}
\text{MS} \Big [ [\text{BDS}]^{(3)}+ \mathcal R^{(3)}(\mu, \epsilon, p_i)  - T_1 - T_2 \Big ] = \int  \Big( -T_3-\Gamma_{11}^{(3)} \Big)~\frac{\text{d}\mu}{\mu} 
 ~,\label{msR}	
\end{equation}
where
\begin{equation}
T_3 := \left (\Gamma_{12}~\frac{\mathcal W_2^{\text{ren}}}{\mathcal W_1^{\text{ren}}}\right )^{(3)}~.\label{deft3}
\end{equation}
For extracting the leading and next-leading divergences of $\mathcal R^{(3)}$, we have to keep track of the leading and next-leading terms in $\log \mu^2$. Anticipating the leading terms to be $\mathcal O(\log^5 \mu^2)$, we will discard all terms of $\mathcal O(\log^3 \mu^2)$ and lower. For example the $[\text{BDS}]^{(3)}$- terms are only $\mathcal O(\log^2 \mu^2)$. Similarly $\int \Gamma_{11} \text{d}\log \mu $ is also $\mathcal O(\log^2 \mu^2)$. 

Therefore, to get the interesting piece of MS$[\mathcal R^{(3)}]$, we only need to calculate 
the terms $T_1$ to $T_3$. Let us discuss them one after the other.

\subsubsection*{Calculating term $T_1$}
Due to the pole in $\mathcal Z_{12}^{(1)}$, see \eqref{z12}, we have to compute 
the term in the large brackets in \eqref{Tdef1} at order $\mathcal O(\epsilon)$. 

Let us start by evaluating $( \mathcal W_1^{\text{ren}(1)}-\mathcal W_2^{\text{ren}(1)} )$. For every cusp we get a term $-\frac{1}{2\epsilon^2}\left( -\mu^2 s_{\text{cusp}} \pm i \varepsilon \right)^{\epsilon}$ at one loop. Minimally subtracting the poles one ends up with
\begin{align}
 \text{MS}\Big[\frac{-1}{2\epsilon^2}\left( -\mu^2 s_{\text{cusp}} \pm i \varepsilon \right)^{\epsilon}\Big] =& -\frac{1}{4}\log ^2 \left(-\mu^2 s_{\text{cusp}} \pm  i \varepsilon \right) -\frac{\epsilon}{12}\log ^3 \left(-\mu^2 s_{\text{cusp}} \pm  i \varepsilon \right) \nonumber\\
 &-\frac{\epsilon^2}{48}\log^4 \left(-\mu^2 s_{\text{cusp}}  \pm i\varepsilon \right) + \mathcal O(\epsilon^3)~.
\end{align}
Collecting all one-loop cusp contributions and using
the factorisation of $ \mathcal W_2$ in the\\ 't Hooft limit one gets
\begin{align}\label{W1mW2}
&\left( \mathcal W_1^{\text{ren}(1)}-\mathcal W_2^{\text{ren}(1)} \right) = \text{MS}\Big[ \frac{1}{2 \epsilon^2} \Big( \left(s \mu^2 x(1-y) \right)^{\epsilon} +\left(s \mu^2 y(1-x) \right)^{\epsilon} +\left(-2p \cdot p_{-} \mu^2 x \right)^{\epsilon}\nonumber\\
&+\left(-2p \cdot p_{+} \mu^2 (1-x) \right)^{\epsilon} +\left(-2q \cdot q_{-} \mu^2 (1-y) \right)^{\epsilon}+\left(-2q \cdot q_{+} \mu^2 y \right)^{\epsilon} \nonumber\\
&-\left(-2p \cdot p_{-} \mu^2  \right)^{\epsilon}-\left(-2p \cdot p_{+} \mu^2 \right)^{\epsilon} -\left(-2q \cdot q_{-} \mu^2 \right)^{\epsilon}-\left(-2q \cdot q_{+} \mu^2\right)^{\epsilon}     \Big) \Big]~.
\end{align}
Above we have suppressed the $i\varepsilon $-prescription. It is the 
same in all terms. Note the absence of the minus sign in the basis 
of the first two $\epsilon$-powers on the r.h.s. It is due to their different
situation concerning the direction of the arrows in fig.\ref{Cross} and \ref{HexaCross}. Later this fact will
be crucial for the generation of different imaginary parts, depending on whether
$s=2pq$ is negative or positive.

The divergent terms contained in the expression for $( \mathcal W_1^{\text{ren}(1)}-\mathcal W_2^{\text{ren}(1)}  )$ are due to  divergences of the BDS structure, present already in a generic configuration, as well as divergences in the one loop ``finite'' part $F(\mu^2,\epsilon,  \{s\})$, which becomes divergent in the self-crossing limit.

Let us now introduce some shorthand notation for the $\epsilon$-expansion
of \eqref{W1mW2}
\begin{equation}
 \left( \mathcal W_1^{\text{ren}(1)}-\mathcal W_2^{\text{ren}(1)} \right) =:  L^{[2]} + L^{[3]} \epsilon +  L^{[4]} \epsilon^2 +\dots ~.
\end{equation}
For the $L^{[k]}$ we find \footnote{We are only interested in the $\mu$-behaviour of these expressions. To handle logarithms of dimensionless quantities  in  the following expressions one should introduce a new scale $\tilde{\mu}$ and write $\log \left( -s \mu^2 \right) = \log \left(-s \tilde{\mu}^2 \right)+\log \frac{\mu^2}{\tilde{\mu}
^2} $. We suppress this negligibility here since $\tilde \mu$ drops out in the end.}
\begin{equation}
L^{[k]} = \frac{1}{k!} \left( \log^k \mu^2 + k \log \left(s  \mathcal X   \pm i\varepsilon \right)  
\cdot  \log^{k-1} \mu^2 \right) + \mathcal O(\log^{k-2} \mu^2) ~ \label{defL},
\end{equation}
using the abbreviation from \eqref{X}.   With this notation one has
\begin{equation}
 \frac{1}{2}\left( \mathcal W_1^{\text{ren}(1)}-\mathcal W_2^{\text{ren}(1)} \right)^2 \Big|_{\mathcal O(\epsilon)}= \epsilon \left(L^{[2]} ~ L^{[3]} \right)~.\label{wsquared}
\end{equation}
Now, using \eqref{defT}, let us turn to 
\begin{align}\label{eq1}
\log \mathcal W_{1}^{\text{ren}(2)} -\log \mathcal W_{2}^{\text{ren}(2)}= & \text{MS}\Big[ [\text{BDS}]^{(2)} + \mathcal R^{(2)} + \mathcal Z_{12}^{(1)}\left( \mathcal W_1^{\text{ren}(1)}-\mathcal W_2^{\text{ren}(1)} \right) \Big]\nonumber\\-&\text{MS}\Big[ [\text{BDS}]_{\text{upper}}^{(2)}+[\text{BDS}]_{\text{lower}}^{(2)} \Big]  -\mathcal R^{(2)}_{\text{upper}}-\mathcal R^{(2)}_{\text{lower}}  ~.
\end{align}
Here some comments are in order. The remainder functions for the upper and lower contours contributing
to $\mathcal W_2$ in the 't Hooft limit, see \eqref{wilson}, do not become divergent in the self-crossing case. Thus they drop their $\mu$ dependence as $\epsilon \rightarrow 0$. For $T_1$ we need \eqref{eq1} at $\mathcal O(\epsilon)$. All BDS terms in \eqref{eq1} will only contribute $\mathcal O(\log^2 \mu^2)$ terms  and the remainder for the upper and lower contours only
 $\mathcal O(\log \mu^2)$ terms. Thus we are left with $\text{MS}[\mathcal   R^{(2)} + \mathcal Z_{12}^{(1)}( \mathcal W_1^{\text{ren}(1)}-\mathcal W_2^{\text{ren}(1)} ) ]$. For the two-loop remainder function we remember the first footnote in the previous section and use  the result from \cite{georgiou} to obtain $\mathcal R^{(2)}= \gamma_{12}^{(1)} \frac{(-\mu^2 s)^{2\epsilon}}{8 \epsilon^3}$. It contributes with $\mathcal O(\log^4 \mu^2)$ at $\mathcal O(\epsilon)$.

Putting this together with  \eqref{z12},\eqref{Tdef1}, \eqref{defL} and\eqref{wsquared} we find 
\begin{align}
T_1 = -\frac{\gamma_{12}^{(1)}}{2}~ L^{[2]} ~ L^{[3]}  + \frac{\big (\gamma_{12}^{(1)}
\big )^2}{24} \log^4 \left(-\mu^2 s \right)- \frac{\big (\gamma_{12}^{(1)}\big )^2}{4} ~ L^{[4]}~ +\mathcal O(\log^3 \mu^2)~.
\end{align}
 
\subsubsection*{Calculating term $T_2$}
At first we calculate the combination of $\mathcal Z$-factors in \eqref{Tdef2} 
\begin{equation}
\big (\mathcal Z_{12}^{(1)} \big )^2 +\mathcal Z_{12}^{(1)} \mathcal Z_{11}^{(1)} - \mathcal Z_{12}^{(2)} = \frac{\gamma_{12}^{(1)}}{4 \epsilon^2}\left ( \gamma_{12}^{(1)} -\log\left(-\mu^2 s  \mathcal X \pm i\varepsilon   \right) \right)  +\frac{\gamma_{12}^{(2)}}{4 \epsilon}-\frac{\gamma_{12}^{(1)}}{8\epsilon^3}~.
\end{equation}
Thus it is obvious that one has to expand $( \mathcal W_1^{\text{ren}(1)}-\mathcal W_2^{\text{ren}(1)} )$ up to $\mathcal O(\epsilon^3)$. The final result for $T_2$ is
\begin{equation}
T_2 = -\frac{\gamma_{12}^{(1)}}{8} ~ L^{[5]} ~+\frac{\gamma_{12}^{(1)}}{4}\left( \gamma_{12}^{(1)} -\log\left(-\mu^2 s \mathcal X   \pm i\varepsilon \right) \right)~ L^{[4]}~+~\mathcal O(\log^3 \mu^2)~.
\end{equation}

\subsubsection*{Calculating term $T_3$}
Here we have to keep terms including $\mathcal O(\log^3 \mu^2)$, since $T_3$ is integrated in \eqref{msR}. From \eqref{deft3} we
get
\beq
T_3~=~\Gamma_{12}^{(1)}\Big (\frac{\mathcal W_2^{\text{ren}}}{\mathcal W_1^{\text{ren}}}\Big )^{(2)}+\Gamma_{12}^{(2)}\Big (\frac{\mathcal W_2^{\text{ren}}}{\mathcal W_1^{\text{ren}}}\Big )^{(1)}+~\Gamma_{12}^{(3)}~.\label{t3det}
\eeq   
Due to the structure of the crossing anomalous dimension matrix in \eqref{gamma} 
we know that the $\Gamma_{12}^{(k)}$ are independent of $\mu$. Furthermore, since 
$\mathcal W_1^{\text{ren}(1)}$ and $\mathcal W_2^{\text{ren}(1)}$ contain at
most  $\mathcal O(\log^2\mu^2)$ terms, we have to keep track of the first term in 
\eqref{t3det} only.

Then for 
\begin{equation}
\left(\frac{\mathcal W_2^{\text{ren}}}{\mathcal W_1^{\text{ren}}}\right)^{(2)} = \log \mathcal W_2^{\text{ren}(2)}-\log \mathcal W_1^{\text{ren}(2)}+\frac{1}{2}\left( \mathcal W_1^{\text{ren}(1)} - \mathcal W_1^{\text{ren}(1)}\right)^2~
\end{equation}
similar arguments as above allow to neglect all BDS terms. Thus we can continue with   
\begin{equation}
\left(\frac{\mathcal W_2^{\text{ren}}}{\mathcal W_1^{\text{ren}}}\right)^{(2)} = - \text{MS}\Big[ \mathcal R^{(2)}  + \mathcal Z_{12}^{(1)}\left( \mathcal W_1^{\text{ren}(1)}-\mathcal W_2^{\text{ren}(1)} \right) \Big]+\frac{1}{2}\left( \mathcal W_1^{\text{ren}(1)} - \mathcal W_1^{\text{ren}(1)}\right)^2~.\nonumber
\end{equation}
Together with \eqref{W1mW2}, \eqref{z12} and  $\Gamma_{12}^{(1)}=\gamma_{12}^{(1)}$, see \eqref{gamma}, we finally get \begin{equation}
T_3 = - \frac{\big (\gamma_{12}^{(1)}\big )^2}{6} \log^3 (-\mu^2 s )+\frac{\big (\gamma_{12}^{(1)}\big )^2}{2}~ L^{[3]} ~ + \frac{\gamma_{12}^{(1)}}{2}\left( L^{[2]} \right)^2~+~\mathcal O(\log^2 \mu^2)~.
\end{equation}
\\

Combining the three terms that we calculated above and performing the  $\text{d}\log \mu$ integration, needed for the third term in \eqref{msR}, we arrive at
\begin{align}
\text{MS} \Big [\mathcal R^{(3)}\Big ]=&-\frac{21}{320}~\gamma_{12}^{(1)}
\log ^5\mu^2 +\frac{5}{96}~\big (\gamma_{12}^{(1)}\big )^2\log ^4\mu^2~\label{aaa}  \\
&-\frac{\gamma_{12}^{(1)}}{192}~\big (2 \log  (-s\mathcal X \pm i\varepsilon ) 
+61 \log (s\mathcal X \pm i\varepsilon )\big )~\log ^4\mu^2~  +\mathcal O(\log^3\mu^2) \nonumber~.
\end{align} 
We know that for $\epsilon\neq 0$ the remainder $\mathcal R(\mu,\epsilon,
\{s\}) $ depends on $\mu$ via $a_{\text{bare}}=a \mu^{2\epsilon}$ only. Thus the source for the $\log^5 \mu^2$ and $\log^4 \mu^2$ term has to be a term like
\begin{equation}
\left(A + \epsilon B \right)\frac{(\mu^2)^{3 \epsilon}}{\epsilon^5} = \frac{A}{\epsilon^5}+\frac{B+3A\log \mu^2}{\epsilon^4} + \dots + \frac{27}{8}B \log^4 \mu^2 +\frac{81}{40}A\log^5 \mu^2~.
\end{equation}
This way one can reconstruct the leading divergences of $\mathcal R^{(3)}$. We also take into account, that the factor with the two $s$-dependent logarithms in the second line of \eqref{aaa} generates  different imaginary parts, depending on the sign of $s=2pq$. Then with $\gamma_{12}^{(1)}$ from \eqref{gamma-korch} we get for $pq<0$
\be
 \mathcal R^{(3)}(\mu,\epsilon,\{s\}) =\pm \frac{7}{108}\pi i~\big (
\epsilon^{-5}+3\log (\vert 2pq\vert \mu^2\mathcal X)~\epsilon^{-4}\big )
-\frac{\pi^2}{4}~\epsilon^{-4}+\mathcal O(\epsilon^{-3})~,
 \ee
and for $pq>0$
\be
 \mathcal R^{(3)}(\mu,\epsilon,\{s\}) =\mp \frac{7}{108}\pi i~\big (
\epsilon^{-5}+3\log (2pq \mu^2\mathcal X)~\epsilon^{-4}\big )
-\frac{\pi^2}{18}~\epsilon^{-4}+\mathcal O(\epsilon^{-3})~.
 \ee
Using the same arguments as for the two-loop remainder in section 3, one 
now can derive information on the behaviour of the (for generic
non-crossing configurations) finite three-loop remainder $\mathcal R(\{u\})$
if it approaches a configuration with crossing edges. We use \eqref{log-u-z},
\eqref{log-u-z-2}
and \eqref{2-loop-trans}, but now with $\alpha$-coefficients for the three loop case from
appendix B, i.e. $\alpha_5=\frac{9}{20},~\alpha_4=\frac{3}{4}$.\\ 
This yields 
\footnote{The following formul\ae \, are derived  for the hexagon approaching the
configuration of fig.\ref{HexaCross}. But of course, they hold for all $n$-gons ($n\geq 6$)
if $u_2$ is replaced by the cross-ratio constructed out of the vertices of the 
asymptotically crossing edges with the same pattern.}
for $pq<0$
\be 
\mathcal R^{(3)}(\{u\})=\mp~\frac{7}{240}\pi i~\log^5(u_2-1)-\frac{3}{16}\pi^2~\log^4(u_2-1)+\mathcal O\left (\log^2(u_2-1)\right )~,
\ee
and for $pq>0$
\be 
\mathcal R^{(3)}(\{u\})=\pm~\frac{7}{240}\pi i~\log^5(1-u_2)-\frac{1}{24}\pi^2~\log^4(1-u_2)+\mathcal O\left (\log^2(1-u_2)\right )~.\label{final}
\ee

\section{Conclusions}
%%%%%%%%%%%%%%%%%%%%%%%%
Using RG-technique we calculated the coefficients of the leading and next-leading
pole terms for the dimensionally regularised three-loop remainder function
of Wilson loops for null $n$-gons in a configuration with two crossing edges.
With a heuristically derived translation rule between coefficients
in dimensional versus point splitting (off self-crossing) regularisation
we were able to convert this into a statement on the singular behaviour of the remainder function in four dimensions. This function depends on cross-ratios only and develops singularities $\propto \log^5|u-1|$ and $\propto \log^4|u-1|$ when the approach to self-crossing enforces  $u\rightarrow 1$ for the characteristic cross-ratio.

Both for the two-loop and three-loop case the coefficient of the next-leading term
in dimensional regularisation is not conformally invariant. The translation {\it has to} generate a conformal invariant coefficient of the next-leading $\log|u-1|$ power. This fixes the relative weight of
the two leading translation factors. The ratio agrees in both cases with the prediction from our heuristic translation rule.

Another independent check comes from our discussion in section 3. We showed
that for the hexagon two-loop remainder the resulting
coefficients agree with those of the singularities of the suitably continued complete function of \cite{goncharov}. 

We have taken care of sign subtleties for certain imaginary parts related to both the sign of the Mandelstam variable characterising the crossing
and the sign of the $i\varepsilon$-prescription in the gluon propagator.
This gives a set of four pairs of coefficients for the leading and next-leading
singularity. In the analytic continuation of the complete two loop remainder out of the Euclidean region they all found their correspondence
by encircling the branch points of polylogarithms in a suitable manner. 

In the three-loop case the access to these four pairs could be helpful
for speculations about that piece of the wanted complete function,
which generates our result in the limit under discussion. Clearly,
free coefficients $\alpha ,\beta,\gamma, \dots$ in $\alpha  \text{Li}_6+\beta 
(\text{Li}_3)^2+\gamma  \text{Li}_2~\text{Li}_4 + \dots$ would allow to fit an arbitrary pair of coefficients in
the asymptotic result. But if one requires the simultaneous fit for all four paths of continuation, only the ratio of coefficients found
in our paper can be fitted. 

The obvious targets for a continuation of this work are the remainder
functions in higher loop orders both in the case of crossing edges
and in the case of coinciding vertices studied in \cite{dw}. The
most difficult part seems to be the control over the contribution of
$\mathcal Z$-matrix
and lower order Wilson loop terms to the higher order logarithm of the renormalised Wilson loop. \\

Finally, we have to comment on a disagreement of our translation rule based 
coefficient for the leading singularity  with recent results on the symbol
of the three-loop remainder \cite{henn}. From there it can be read of to be 
\footnote{private communication by L. Dixon and J. Henn} 
$\pm 1/40~\pi i\log^5(1-u_2)$, 
implying a discrepancy by a factor $6/7$ in comparison to our 
equation \eqref{final}. After submitting our first preprint version,
the authors of \cite{caron2}, among other things, independently confirmed 
the results of \cite{henn} 
for the symbol of the hexagon three-loop remainder. 
Therefore, we have to suspect a breakdown of the translation
rule of appendix B starting at three loops.

To get some diagrammatic understanding of both the perfect match at two
loops and the possible breakdown at three loops, we reproduced the leading
divergence of the two-loop remainder in dimensional regularisation by direct
Feynman diagram analysis. For this purpose one has to take into account only 
diagrams with two gluon propagators, ending on the crossing edges and corresponding
to a colour diagram with crossing propagators. Their weight has to be chosen 
according to the non-Abelian exponentiation theorem \cite{taylor}. 

It is interesting to comment on the irrelevance of diagrams containing the three
gluon vertex or propagator corrections. In chapter 2.2 of the third reference 
of \cite{drummond} one finds an illuminating discussion of the mechanism ensuring
for generic (non-crossing) configurations that, in agreement
with the BDS structure, only poles up to order $1/\epsilon ^2$ appear in the
logarithm of the Wilson loop. Single contributing diagrams have poles up to
$1/\epsilon^4$, but both the $1/\epsilon^4$ and $1/\epsilon ^3$ terms cancel between
diagrams with a three gluon vertex, with propagator corrections and with the above 
mentioned  two propagators. 

For the contribution from two
crossing edges this mechanism is no longer valid. As shown in \cite{korchemsky2},
the three gluon vertex diagrams with one gluon on one of the crossing edge
and the other two on the second crossing edge, in all possible positions
relative to the crossing point, sum to zero. Furthermore, the diagrams with
self energy corrections do not contribute to $1/\epsilon^3$, since the 
integrations over the endpoints of the radiative corrected gluon propagator
extend over both sides of the crossing edges. Altogether for the leading divergence
only the diagrams with two propagators contribute. 
The $1/\epsilon ^4$ divergence cancels among them, but just the correct 
$1/\epsilon^3$ term survives. 

Seemingly this reduction to iterated one-loop diagrams is responsible for
the validity of the translation rule up to two loops.    
 
For the three-loop case a similar analysis of the contribution via
non-Abelian exponentiation of the diagrams with three gluon propagators
does not give the correct factor of the $1/\epsilon ^5$ term. This indicates that 
now diagrams with gluon vertices and/or propagator corrections contribute.
This could be the reason for a breakdown of the naive translation rule. 

Further work should give a better understanding of this issue. 
It would be also very interesting to circumvent the translation problem
by formulating the RG technique from the very beginning for a point
splitting regularisation.   
\\[10mm]
\noindent
{\bf Acknowledgement}\\[2mm]
We  thank  George Georgiou, Johannes Henn, George Jorjadze, Chrysostomos Kalou\-sios, Christoph Sieg and Gabriele Travaglini for useful
discussions. This work was supported by DFG via GK 1504 and SFB 647 and by VolkswagenStiftung via grant I/84600.

\section*{Appendix A}
\begin{figure}[h!]
 \centering
 \includegraphics[width=6cm]{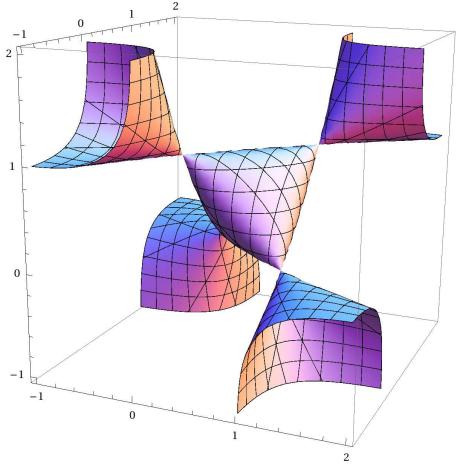} 
\caption{{\it The allowed region for the three cross-ratios of null hexagons in four-dimensional Minkowski space. Points outside the central bag and the four ears are accessible only with null hexagons in $\mathbb R^{(2,2)}$.}}
 \label{bag}
\end{figure}

The range of the three cross-ratios $u_1,u_2,u_3$, which can be realized
with real null hexagons in four-dimensional Minkowski space, has been partly
discussed in ref. \cite{bubble}. Their discussion is based on the analysis
of the remaining freedom after $x_5,x_6,x_1$ have been sent to 
infinity by conformal maps, but restricted to the case with space-like 
$(x_2-x_4)^2$. Including also the case of time-like $(x_2-x_4)^2$ one gets the 
following description of the complete allowed range
\bea
U_{13}:=u_1u_3~(u_1-1)(u_3-1)&\geq &0~,\\
u_1u_3+(u_1-1)(u_3-1)-2\sqrt{U_{13}}&\leq &u_2~\leq~u_1u_3+(u_1-1)(u_3-1)+2\sqrt{U_{13}}~.\nonumber\label{u-range}
\eea
A corresponding graph is shown in fig.\ref{bag}. Allowed are all points inside
the central bag and inside the four ears extending up to infinity. The other regions
of $(u_1,u_2,u_3)$-space are accessible with real configurations in $\mathbb R^{(2,2)}$ only. A symmetric characterisation of the surface separating the $\mathbb R^{(3,1)}$ and $\mathbb R^{(2,2)}$ is given by  \cite{bubble}
\beq
4u_1u_2u_3~=~(1-u_1-u_2-u_3)^2~.\label{u-symm}
\eeq
%%%%%%%%%%%%%%%%%%%%%
%%%%%%%%%%%%%%%%%%%%%%%%%%%%%%%%%%%%%

In null hexagons with a self-crossing necessarily one of the $u_i$ equals one and
the remaining two are equal to each other \cite{georgiou}. 
In the main text we have already noticed that a self-crossing situation with
$pq>0$ can be reached only via $u_i\rightarrow 1-0$, which means from inside the 
central bag in fig.\ref{bag}. If $pq<0$ one can approach a self-crossing configuration
only from inside the ears in fig.\ref{bag}.\\

As explained in the main text, the three cross-ratios do not uniquely fix
the conformal class of the hexagon. While all self-crossing configurations
have two equal cross-ratios and one equal to one, not all configurations with
such a cross-ratio pattern are self-crossing. Clearly, self-crossing and not self-crossing
configurations cannot be conformally mapped to each other. 

\begin{figure}[h!]
 \centering
 \includegraphics[width=8cm]{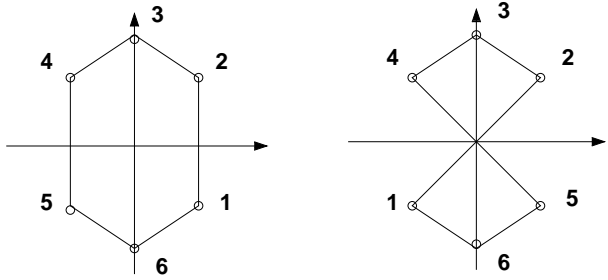} 
\caption{{\it Projection to the (1,2)-plane of a null hexagon without self-crossing, but with $u_2=1$ (left) and after rotating $x_1$ and $x_5$ with angle $\pi$ in the (1,3)-plane (right).}}
 \label{def}
\end{figure}

To clarify how
self-crossing configurations with diverging remainder function can be reached
by continuous deformation from configurations without self-crossing, but with
say $u_2=1,~~u=u_1=u_3$, we consider
\bea
x_2&=&(2,1,1,0),~~x_3~=~(2-\sqrt{b^2-2b+2},0,b,0),~~x_4~=~(2,-1,1,0)\\
x_1&=&(x_{10},\cos\psi,-1,\sin\psi),~~x_5=(x_{50},-\cos\psi,-1,-\sin\psi),~~x_6=(x_{60},0,-b,0),\nonumber \label{deform}
\eea
with
\beq
x_{10}~=~x_{50}~=~2-\sqrt{6-2  \cos\psi},~~x_{60}~=~2-\sqrt{6-2\cos\psi}+\sqrt{2+b^2-2b}~.\label{deform2}
\eeq
The related cross-ratios are
\bea
u_2&=&\frac{1}{\cos ^2\psi}~,\nonumber\\
u&=&-~\frac{(~3-\cos\psi-2b-\sqrt{6-2\cos\psi}\sqrt{b^2-2b+2}~)^{2}}{\cos\psi~(14-8b-2\cos\psi-4\sqrt{6-2\cos\psi}\sqrt{b^2-2b+2})}~.\label{u-interpol}
\eea
For $\psi=0$ we have a not self-crossing hexagon with $u_2=1$ and for $\psi=\pi$
a self-crossing hexagon with again $u_2=1$. Their projections on the $(1,2)$-plane
are shown in fig.\ref{def}. 
%%%%%%%%%%%%%%%%%%%%%
%%%%%%%%%%%%%%%%%%%%%%%%%%%%%%%%%%%%%    

The cross-ratio $u_2$ goes through infinity. Therefore, to follow the path of the 
cross-ratios for all $\psi\in(0,\pi)$, it is more  convenient to consider its
description in terms of $1/u$ and $1/u_2$. Then \eqref{u-interpol} results in a 
continuous and finite path in the $(1/u,1/u_2)$-plane. Along the path $1/u_2$
goes from 1 to 0 and back to 1. For generic $b$ the initial and final values of
$1/u$ differ, for the special case $b=-\frac{2(1+2\sqrt{2})}{7}=-1.094$ they agree.

The source for divergences in the analytic result of \cite{goncharov}
are polylogarithms with argument $(1-\frac{1}{u_i})$. Now $(1-\frac{1}{u_2})$
for $\psi=0$ to $\psi=\pi$ goes from zero to one and back to zero. For the behaviour
near one, where the polylogarithms have a branch point with a cut extending to $+\infty$, we have to remember that the $i\varepsilon$-prescription of the gluon
propagator will take care for encircling the branch point. Let us handle the
$(x_j-x_k)^2$ factors in the cross-ratios in the same manner as the squared distances
in the gluon propagator. Then we find, that
the argument of the polylogarithms encircles the branch point at one 
in a counterclockwise (clockwise) manner for the standard (switched) $i\varepsilon$-prescription .
%%%%%%%%%%%%%%%%%%%%%%%%%%%%%
\section*{Appendix B}
\noindent
Renormalisation $Z$-factors 
are given as formal power series with terms $g^{2l}\epsilon^{-m}$ for
dimensional regularisation and with terms $g^{2l}\log^m(\frac{1}{-z^2_{\perp}\mu^2})$ for
a regularisation with a position space cutoff $z$ ($\mu$ RG-scale).
In the case of correlation functions of local operators only terms
with $m\leq l$ contribute. The complete all loop information on
$\beta$-functions and anomalous dimensions is contained in the coefficients
of the $m=1$ terms. On the other side, the coefficients of the terms
with $m=l$ are fixed by one loop information and are independent of the
renormalisation scheme.  

In contrast to this situation, for polygonal Wilson loops with light-like 
edges also terms with $m>l$
appear. In dimensional regularisation the RG-scale enters exclusively 
in the combination $g^2\mu ^{2\epsilon}$ and we have to deal with ($s$ some
squared distance of polygon vertices)
\beq
g^{2l}~\frac{1}{\epsilon ^m}(-\mu ^2 s)^{l\epsilon}~=~g^{2l}~\frac{1}{\epsilon ^m}
\left (1+\cdots
+\frac{(l\epsilon)^{m-l}}{(m-l)!}~\log^{m-l}(-\mu^2s)+\cdots\right )~.\label{epsnm}
\eeq
The analogous term in cutoff regularisation looks like
\begin{align}
g^{2l}~\log^{m}(\frac{s}{z_{\perp}^2})&=g^{2l}~\Big (\log(-\mu^2s)+\log(\frac{1}{-\mu^2z_{\perp}^2})
\Big )^m\nonumber\\
&= \cdots  +g^{2l}~{m\choose l}\log^{m-l}(-\mu^2s)~\log^{l}\left(\frac{1}{-\mu^2z_{\perp}^2}\right)~+~\cdots \label{cutoffnm}
\end{align}
In both cases a term with $m>l$ generates descendents with lower powers of $\frac{1}{\epsilon}$ or $\log \frac{1}{-\mu^2z_{\perp}^2}$, respectively. 
\\*Now we assume the existence of some translation factor $\alpha$
\beq
g^{2l}~\frac{1}{\epsilon ^m}(-\mu ^2 s)^{l\epsilon}~\Leftrightarrow ~\alpha ~g^{2l}~\log^{m}(\frac{s}{z_{\perp}^2})~.
\eeq
Inserting in this correspondence \eqref{epsnm} and \eqref{cutoffnm} one
finds, by comparison of the coefficients of $\log^{m-l}(-\mu^2s)$
\beq
g^{2l}\frac{1}{\epsilon^l}~\frac{l^{m-l}}{(m-l)!}~\Leftrightarrow ~\alpha ~g^{2l}
\log^{l}\left(\frac{1}{-\mu^2z_{\perp}^2}\right){m\choose l}~.\label{adjust}
\eeq
\\*Motivated by the scheme independence (for local operators) of factors in front of terms with equal powers of $g^2$ and $1/\epsilon$ or $\log (\frac{1}{-\mu^2z_{\perp}^2})$, we further {\it assume} that the translation 
factor $\alpha$ has to be chosen in such a manner that \eqref{adjust}
becomes $g^{2l}/\epsilon^l\Leftrightarrow g^{2l}\log ^l$. This fixes
\beq
\alpha~=~\frac{l^{m-l}~l!}{m!}~.\label{trans-rule}
\eeq
Looking then on the coefficients of $(\log(-\mu^2s))^0$ we
get
\beq
g^{2l}~\frac{1}{\epsilon ^m}~\Leftrightarrow ~\alpha ~g^{2l}~\log^{m}\left(\frac{1}{
-\mu^2z_{\perp}^2}\right)~, 
\eeq
with $\alpha$ from \eqref{trans-rule}.
\\*\\*	In particular this means\\[2mm]
- at one loop~~~~: $1/\epsilon ^2~\Leftrightarrow ~1/2~ \log^2~,~~1/\epsilon ~\Leftrightarrow ~ \log$\\
- at two loops~~~: $1/\epsilon ^3~\Leftrightarrow ~2/3~\log^3~,~~1/\epsilon ^2~\Leftrightarrow ~\log^2$\\
- at three loops~: $1/\epsilon ^5~\Leftrightarrow ~9/20~\log^5~,~~1/\epsilon ^4~\Leftrightarrow ~3/4~\log^4$\\[2mm]
It is straightforward to prove at least the first statement by an explicit 
calculation. Two further independent checks are emphasized in the conclusions.

\newpage
 
%%%%%%%%%%%%%%%%%%%%%%%%%%%%%%%%%%


\begin{thebibliography}{99}

\bibitem{alday-malda}
  L.~F.~Alday and J.~M.~Maldacena,
  %``Gluon scattering amplitudes at strong coupling,''
  JHEP {\bf 0706} (2007) 064
  [arXiv:0705.0303 [hep-th]].
  %%CITATION = JHEPA,0706,064;%%
$\bullet$ 
L.~F.~Alday and J.~Maldacena,
  %``Comments on gluon scattering amplitudes via AdS/CFT,''
  JHEP {\bf 0711} (2007) 068
  [arXiv:0710.1060 [hep-th]].
  %%CITATION = JHEPA,0711,068;%%

\bibitem{drummond}
  G.~P.~Korchemsky, J.~M.~Drummond and E.~Sokatchev,
  %``Conformal properties of four-gluon planar amplitudes and Wilson loops,''
  Nucl.\ Phys.\ B {\bf 795} (2008) 385
  [arXiv:0707.0243 [hep-th]].
  %%CITATION = ARXIV:0707.0243;%%
  $\bullet$
  A.~Brandhuber, P.~Heslop and G.~Travaglini,
  %``MHV amplitudes in N=4 super Yang-Mills and Wilson loops,''
  Nucl.\ Phys.\ B {\bf 794} (2008) 231
  [arXiv:0707.1153 [hep-th]].
  %%CITATION = ARXIV:0707.1153;%%
  $\bullet$
  J.~M.~Drummond, J.~Henn, G.~P.~Korchemsky and E.~Sokatchev,
  %``On planar gluon amplitudes/Wilson loops duality,''
  Nucl.\ Phys.\  B {\bf 795} (2008) 52
  [arXiv:0709.2368 [hep-th]].
  %%CITATION = NUPHA,B795,52;%%
  $\bullet$
  J.~M.~Drummond, J.~Henn, G.~P.~Korchemsky and E.~Sokatchev,
  %``Hexagon Wilson loop = six-gluon MHV amplitude,''
  Nucl.\ Phys.\  B {\bf 815} (2009) 142
  [arXiv:0803.1466 [hep-th]].
  %%CITATION = NUPHA,B815,142;%%

\bibitem{bds}
  C.~Anastasiou, Z.~Bern, L.~J.~Dixon and D.~A.~Kosower,
  %``Planar amplitudes in maximally supersymmetric Yang-Mills theory,''
  Phys.\ Rev.\ Lett.\  {\bf 91} (2003) 251602
  [arXiv:hep-th/0309040].
  %%CITATION = PRLTA,91,251602;%%
  $\bullet$ 
    Z.~Bern, L.~J.~Dixon and V.~A.~Smirnov,
  %``Iteration of planar amplitudes in maximally supersymmetric Yang-Mills
  %theory at three loops and beyond,''
  Phys.\ Rev.\  D {\bf 72} (2005) 085001
  [arXiv:hep-th/0505205].
  %%CITATION = PHRVA,D72,085001;%%

\bibitem{duhr}
  V.~Del Duca, C.~Duhr and V.~A.~Smirnov,
  %``The Two-Loop Hexagon Wilson Loop in N = 4 SYM,''
  JHEP {\bf 1005} (2010) 084
  [arXiv:1003.1702 [hep-th]].
  %%CITATION = JHEPA,1005,084;%%

\bibitem{goncharov}
A.~B.~Goncharov, M.~Spradlin, C.~Vergu and A.~Volovich,
  %``Classical Polylogarithms for Amplitudes and Wilson Loops,''
  Phys.\ Rev.\ Lett.\  {\bf 105} (2010) 151605
  [arXiv:1006.5703 [hep-th]].
  %%CITATION = PRLTA,105,151605;%%

\bibitem{duhr-8}
   V.~Del Duca, C.~Duhr and V.~A.~Smirnov,
  %``A Two-Loop Octagon Wilson Loop in N = 4 SYM,''
  JHEP {\bf 1009} (2010) 015
  [arXiv:1006.4127 [hep-th]].
  %%CITATION = JHEPA,1009,015;%%

\bibitem{caron}
  S.~Caron-Huot,
  %``Superconformal symmetry and two-loop amplitudes in planar N=4 super 
  % Yang-Mills,''
  JHEP {\bf 1112} (2011) 066
  [arXiv:1105.5606 [hep-th]].
  %%CITATION = ARXIV:1105.5606;%%  

\bibitem{henn}
  L.~J.~Dixon, J.~M.~Drummond and J.~M.~Henn,
  %``Bootstrapping the three-loop hexagon,''
  JHEP {\bf 1111} (2011) 023
  [arXiv:1108.4461 [hep-th]].
  %%CITATION = ARXIV:1108.4461;%%

\bibitem{georgiou}
  G.~Georgiou,
  %``Null Wilson loops with a self-crossing and the Wilson loop/amplitude 
  % conjecture,''
  JHEP {\bf 0909 } (2009)  021
  [arXiv:0904.4675 [hep-th]].
  %%CITATION = JHEPA,0909,021;%%

\bibitem{dw}
H.~Dorn and S.~Wuttke,
  %``Wilson loop remainder function for null polygons in the limit of
  %self-crossing,''
  JHEP {\bf 1105} (2011) 114
  [arXiv:1104.2469 [hep-th]].
  %%CITATION = JHEPA,1105,114;%%

\bibitem{pol}
  A.~M.~Polyakov,
  %``Gauge Fields As Rings Of Glue,''
  Nucl.\ Phys.\  B {\bf 164} (1980) 171.
  %%CITATION = NUPHA,B164,171;%%

\bibitem{brandt}
  R.~A.~Brandt, F.~Neri and M.~A.~Sato,
  %``Renormalization Of Loop Functions For All Loops,''
  Phys.\ Rev.\  D {\bf 24} (1981) 879.\\
  %%CITATION = PHRVA,D24,879;%%
  $\bullet $
  R.~A.~Brandt, A.~Gocksch, M.~A.~Sato and F.~Neri,
  %``Loop Space,''
  Phys.\ Rev.\  D {\bf 26} (1982) 3611.
  %%CITATION = PHRVA,D26,3611;%%

\bibitem{dorn}
  H.~Dorn,
  %``RENORMALIZATION OF PATH ORDERED PHASE FACTORS AND RELATED HADRON OPERATORS
  %IN GAUGE FIELD THEORIES,''
  Fortsch.\ Phys.\  {\bf 34} (1986) 11.
  %%CITATION = FPYKA,34,11;%%

\bibitem{korchemsky2}
  I.~A.~Korchemskaya and G.~P.~Korchemsky,
  %``High-energy scattering in QCD and cross singularities of Wilson loops,''
  Nucl.\ Phys.\  B {\bf 437} (1995) 127
  [arXiv:hep-ph/9409446].
  %%CITATION = NUPHA,B437,127;%%

\bibitem{korchrad}
  G.~P.~Korchemsky and A.~V.~Radyushkin,
  %``Renormalization of the Wilson Loops Beyond the Leading Order,''
  Nucl.\ Phys.\  B {\bf 283} (1987) 342.
  %%CITATION = NUPHA,B283,342;%%

\bibitem{korchemsky}
  I.~A.~Korchemskaya and G.~P.~Korchemsky,
  %``On lightlike Wilson loops,''
  Phys.\ Lett.\  B {\bf 287}, 169 (1992).
  %%CITATION = PHLTA,B287,169;%%


\bibitem{brandhuber10}
  A.~Brandhuber, P.~Heslop, P.~Katsaroumpas, D.~Nguyen, B.~Spence, M.~Spradlin and G.~Travaglini,
  %``A Surprise in the Amplitude/Wilson Loop Duality,''
  JHEP {\bf 1007} (2010) 080
  [arXiv:1004.2855 [hep-th]].
  %%CITATION = JHEPA,1007,080;%%


\bibitem{bubble}
L.~F.~Alday, D.~Gaiotto and J.~Maldacena,
  %``Thermodynamic Bubble Ansatz,''
  JHEP {\bf 1109 } (2011)  032.
  [arXiv:0911.4708 [hep-th]].
  %%CITATION = ARXIV:0911.4708;%%

  \bibitem{caron2}
  S.~Caron-Huot and S.~He,
  ``Jumpstarting the all-loop S-matrix of planar N=4 super Yang-Mills,''
  arXiv:1112.1060 [hep-th].
  %%CITATION = ARXIV:1112.1060;%%

 \bibitem{taylor}
  J.~G.~M.~Gatheral,
  %``Exponentiation Of Eikonal Cross-sections In Nonabelian Gauge Theories,''
  Phys.\ Lett.\ B {\bf 133} (1983) 90.
  %%CITATION = PHLTA,B133,90;%%
  $\bullet$ 
  J.~Frenkel and J.~C.~Taylor,
  %``Non-Abelian Eikonal Exponentiation,''
  Nucl.\ Phys.\ B {\bf 246} (1984) 231.
  %%CITATION = NUPHA,B246,231;%%

\end{thebibliography}
\end{document}